\documentclass{ifacconf}

\usepackage{graphicx}      
\usepackage{natbib}        

\usepackage{amsmath}
\usepackage{amssymb}
\usepackage{siunitx}
\DeclareMathOperator*{\arginf}{arg\,inf}
\begin{document}
\begin{frontmatter}
\title{Learning Data-Driven PCHD Models for Control Engineering Applications \thanksref{footnoteinfo}} 

\thanks[footnoteinfo]{This work was developed in the junior research group DART (Daten\-ge\-trie\-be\-ne Methoden in der Regelungstechnik), Paderborn University, and funded by the Federal Ministry of Education and Research of Germany (BMBF - Bundes\-ministerium für Bildung und Forschung) under the funding code 01IS20052. The responsibility for the content of this publication lies with the authors.\\
© 2022 the authors. \textit{This work has been accepted to IFAC for publication under a Creative Commons Licence CC-BY-NC-ND}.}

\author{Annika Junker,}
\author{Julia Timmermann,}
\author{Ansgar Trächtler}

\address{Heinz Nixdorf Institute, Paderborn University, Germany (e-mail: \{annika.junker, julia.timmermann, ansgar.traechtler\}@hni.upb.de)}

\begin{abstract}                
The design of control engineering applications usually requires a model that accurately represents the dynamics of the real system. In addition to classical physical modeling, powerful data-driven approaches are increasingly used. However, the resulting models are not necessarily in a form that is advantageous for controller design. In the control engineering domain, it is highly beneficial if the system dynamics is given in PCHD form (Port-Controlled Hamiltonian Systems with Dissipation) because globally stable control laws can be easily realized while physical interpretability is guaranteed. In this work, we exploit the advantages of both strategies and present a new framework to obtain nonlinear high accurate system models in a data-driven way that are directly in PCHD form. We demonstrate the success of our method by model-based application on an academic example, as well as experimentally on a test bed.  
\end{abstract}


\end{frontmatter}
\section{Introduction}\label{sec:introduction}
In the modeling of technical systems, data-driven methods such as neural networks have been increasingly used in recent years. Compared to classical physically-based modeling of technical systems, data-driven non-parametric approaches enable the representation of generic correlations without being constrained to a given parametric model. In most cases, data-driven model building results in a model that accurately represents the system dynamics but is no longer physically interpretable.

Several popular and powerful numerical methods for system identification have been developed in recent years within Koopman operator theory (\cite{Sch10,PBK16,BPK16,KM18b}). These methods extract a dynamic system from the measured data of the underlying system by \textit{lifting} the states into a generally higher-dimensional space and approximating the dynamics as a linear system, which is called \textit{Extended Dynamic Mode Decomposition (EDMD)} (\cite{WKR15}). The characteristics of a linear system description open up new possibilities for applications, e.\,g., gaining deeper insight into the system by analyzing system-theoretic properties (\cite{JTT22}). However, it is not guaranteed that the linearly approximated model correctly represents the system theoretic properties of the underlying nonlinear system. Currently, some approaches impose stability properties on the data-driven model by forcing the eigenvalues of the Koopman operator to be stable (\cite{MAM20}). Our approach to ensuring physically plausible models is based on the property of passivity. 
 
\textit{Passive systems}, or specifically \textit{Port-Controlled Hamiltonian systems with Dissipation (PCHD)} exhibit highly useful properties for the controller design (\cite{BIW91}) since they are physically plausible and easy to interpret. Moreover, they can be used to straightforwardly obtain globally stable control laws, because a negative feedback loop consisting of two passive systems is passive (\cite{SJK97}) and there is a strong connection to Lyapunov stability (\cite{Kha15}). However, a complete nonlinear system model is required to analytically obtain such a PCHD form. Therefore, we propose a new framework to directly learn such PCHD models in a data-driven way. Inspired by the system identification methods of the Koopman operator, we follow the strategy of a hybrid approach, i.\,e., we use measurement data of the original system combined with prior physical knowledge about the energy stored in the system. 

The work is structured as follows: Section\,\ref{sec:background} reviews the basics of passive systems, PCHD models, Extended Dynamic Mode Decomposition, and stable Koopman operators. In Section\,\ref{sec:algorithm}, we introduce the proposed algorithm for obtaining a data-driven model in PCHD form. Section\,\ref{sec:results} demonstrates the success of the framework with simulated and experimental results. Section\,\ref{sec:conclusion} summarizes the approach and reveals possible future research.

\textbf{Notation:} Assume $\vec{A}\in\mathbb{R}^{n\times n}$. $\vec{A}^\top$ denotes the transpose and $\vec{A}^+$ the Moore-Penrose inverse of $\vec{A}$. $\vec{A}$ is said to be positive-definite ($\succ$) if $\vec{x}^\top\vec{A}\vec{x}>0$ for all $\vec{x}\in\mathbb{R}^n \setminus{\vec{0}}$ and said to be positive semi-definite ($\succeq$) if $\vec{x}^\top\vec{A}\vec{x}\geq0$ for all $\vec{x}\in\mathbb{R}^n$. 
$\lVert\vec{A}\rVert$ denotes the spectral norm and  $\lVert\vec{A}\rVert_F$ the Frobenius norm of $\vec{A}$. $\mathcal{O}(n)$ is the group of $n\times n$ orthogonal matrices. A real-valued, continuously differentiable function $f$ is called positive-definite in a neighborhood $D$ of the origin if $f(\vec{0})=0$ and $f(\vec{x})>0$ for $\vec{x}\neq\vec{0}$.
In all cases, the index $t$ denotes discrete-time system descriptions. 
\section{Background}\label{sec:background}
In the following, we introduce the notion of passivity\,(\ref{subsec:passive_systems}) and discuss PCHD systems\,(\ref{subsec:pchd}) in more detail. Moreover, we review the numerical system identification method \textit{Extended Dynamic Mode Decomposition for control}\,(\ref{subsec:edmdc}) and raise the topic of stable Koopman operators\,(\ref{subsec:stable_KO}).
\subsection{Passive Systems}\label{subsec:passive_systems}
The notion of passivity was motivated by energy dissipation of a dynamical system and has been used to analyze the stability of a general class of interconnected nonlinear systems (\cite{BIW91}). Hyperstability is closely related to passivity and refers to linear systems that can be described by a transfer function that is positive real  (\cite{And68,Pop63,Pop73}). \cite{BIW91} established the concept of passivity for nonlinear systems. Passive systems are always stable and the concept can be used to asymptotically stabilize nonlinear feedback systems,  making such a system description highly desirable. 

Consider continuous-time nonlinear state-space models
\begin{subequations}\label{eq:nl_sys}
	\begin{align}
		\dot{\vec{x}}&=\vec{f}(\vec{x},\vec{u}),\\
		\vec{y}&=\vec{g}(\vec{x},\vec{u}),
	\end{align}
\end{subequations}
where the vector $\vec{x}\in\mathbb{R}^n$ denotes the states of the system and $\dot{\vec{x}}\in\mathbb{R}^n$ the time derivative of the states. $\vec{u}\in\mathbb{R}^p$ denotes the system inputs and ${\vec{y}\in\mathbb{R}^q}$ the system outputs. $\vec{f}$ is assumed to be locally Lipschitz, $\vec{g}$ to be continuous, and ${\vec{f}(\vec{0},\vec{0})=\vec{g}(\vec{0},\vec{0})=\vec{0}}$ to be the fixed point.  

The system\,\eqref{eq:nl_sys} with $p=q$ is passive if there exists a continuously differentiable positive semi-definite \textit{storage function} $V:\mathbb{R}^n\rightarrow\mathbb{R}$, such that
\begin{equation}\label{eq:passive_1}
	V(\vec{x}(t))-V(\vec{x}(0))\leq\int_{0}^{t}{\vec{u}^\top(\tau)\vec{y}(\tau)d\tau},
\end{equation}
\begin{equation}\label{eq:passive_2}
	\Rightarrow\dot{V}(\vec{x}(t))=\dfrac{\partial V}{\partial \vec{x}}\dot{\vec{x}}\leq\vec{u}^\top\vec{y}
\end{equation}
for all $(\vec{x},\vec{u})$.
Illustratively, it means
\begin{equation}\label{eq:passive_3}
	\text{stored energy}\leq\text{supplied energy}.
\end{equation}
Moreover, special cases cover strictly passive or lossless systems, where the inequality in\,\eqref{eq:passive_1}-\eqref{eq:passive_3} is replaced by $<$ or $=$, respectively. 

If the system\,\eqref{eq:nl_sys} is passive with a positive-definite storage function $V(\vec{x})$, then $\vec{x}=\vec{0}$ is stable. Furthermore, if the storage function is radially unbounded, the origin will be globally asymptotically stable (\cite{Kha15}).

\subsection{Port-Controlled Hamiltonian Systems with Dissipation}\label{subsec:pchd}
The dynamics of non-resistive physical systems can be given an intrinsic Hamiltonian formulation, leading to \textit{Port-Controlled Hamiltonian (PCH) systems} (\cite{Mv92}) and have been extended by dissipation effects, leading to PCHD (\cite{MOv00})
\begin{subequations}\label{eq:pchd}
	\begin{align}
		\dot{\vec{x}}&=\left(\vec{J}(\vec{x})-\vec{D}(\vec{x})\right)\left(\dfrac{\partial V}{\partial\vec{x}}\right)^\top+\vec{B}(\vec{x})\vec{u},\\
		\vec{y}&=\vec{B}^\top(\vec{x})\left(\dfrac{\partial V}{\partial \vec{x}}\right)^\top,
	\end{align}
\end{subequations}
where $\vec{x}\in\mathbb{R}^n$ contains the states by which the energy is defined and $\vec{u},\vec{y}\in\mathbb{R}^p$ are the port power variables. ${V:\mathbb{R}^n\rightarrow\mathbb{R}}$ is a positive-definite smooth function representing the stored energy. ${\vec{J}(\vec{x})\in\mathbb{R}^{n\times n}}$ is a skew-symmetric matrix defining the energy flow inside the system and ${\vec{D}(\vec{x})\in\mathbb{R}^{n\times n}}$ is a symmetric positive semi-definite matrix defining the energy dissipation effects.

The time derivative of $V$ yields
\begin{align}
	\dot{V}(\vec{x})=\vec{u}^\top\vec{y}-\dfrac{\partial V}{\partial\vec{x}}\vec{D}(\vec{x})\left(\dfrac{\partial V}{\partial\vec{x}}\right)^\top\leq\vec{u}^\top\vec{y},
\end{align}
so that \eqref{eq:pchd} meets the passivity criterion\,\eqref{eq:passive_2}.

\subsection{Extended Dynamic Mode Decomposition with Control}\label{subsec:edmdc}
With the linear but infinite-dimensional Koopman operator (\cite{Koo31}), the dynamics of nonlinear systems can be described linearly by \textit{lifting} the states into a higher-dimensional space (\cite{BBPK16}). For practical applications, the Koopman operator is usually approximated numerically as a finite-dimensional matrix, using the well-known method \textit{Extended Dynamic Mode Decomposition with Control (\mbox{EDMDc})} (\cite{PBK16,WKR15}). Below is a brief description of the algorithm; a detailed utilization of the method illustrated with examples is given in \cite{JTT22}. 

In the following, continuous-time control-affine systems
	\begin{equation}
		\dot{\vec{x}}=\vec{f}(\vec{x})+\vec{B}\vec{u}
		\label{eq:control-affine_system}
	\end{equation}
with a constant input matrix $\vec{B}\in\mathbb{R}^{n\times p}$ are considered, which is a justified restriction in many control engineering applications.

For the approximation of the Koopman operator, $N$ observable functions $\vec{\Psi}(\vec{x})=\begin{bmatrix}
		\psi_1(\vec{x}),\psi_2(\vec{x}),\cdots,\psi_N(\vec{x})
	\end{bmatrix}^\top$ are defined, which lift the states into the higher-dimensional space. The algorithm approximates the dynamics of the lifted states $\vec{\Psi(x)}$ as a discrete-time system
\begin{equation}
	\vec{\Psi}(\vec{x}_{k+1})\approx\vec{K}_t\vec{\Psi}(\vec{x}_k)+\vec{B}_t\vec{u}_k=\begin{bmatrix}
		\vec{K}_t,\vec{B}_t
	\end{bmatrix}\begin{bmatrix}
		\vec{\Psi}(\vec{x}_k)\\ \vec{u}_k
	\end{bmatrix}.
\label{eq:edmd}
\end{equation}
With measurement data
\begin{subequations}\label{eq:data_edmdc}
	\begin{align}
	\label{eq:snapshots_x_edmdc}
	\vec{X}&=\begin{bmatrix}
		\vec{x}_1, \vec{x}_2,\cdots,\vec{x}_{M-1}
	\end{bmatrix}\in \mathbb{R}^{n\times (M-1)},\\
	\vec{X}'&=\begin{bmatrix}
		\vec{x}_2, \vec{x}_3,\cdots,\vec{x}_M
	\end{bmatrix}\in \mathbb{R}^{n\times (M-1)},\\
	\vec{U}&=\begin{bmatrix}
		\vec{u}_1,\vec{u}_2\cdots,\vec{u}_{M-1}
	\end{bmatrix}\in\mathbb{R}^{p\times(M-1)}
\end{align}
\end{subequations}
and
\begin{subequations}
	\begin{align}
		\label{eq:snapshots_psi_edmdc}
		\vec{\Psi}(\vec{X})&=\begin{bmatrix}
			\vec{\Psi}(\vec{x}_1),\cdots,\vec{\Psi}(\vec{x}_{M-1})	
		\end{bmatrix}\in\mathbb{R}^{N\times (M-1)},\\
		\vec{\Psi}(\vec{X'})&=\begin{bmatrix}
			\vec{\Psi}(\vec{x}_2),\cdots,\vec{\Psi}(\vec{x}_M)	
		\end{bmatrix}\in\mathbb{R}^{N\times (M-1)}
	\end{align}
\end{subequations}
results
\begin{align}
	\vec{\Psi(X')}&\approx\vec{K}_t\vec{\Psi(X)}+\vec{B}_t\vec{U}
	=\begin{bmatrix}
		\vec{K}_t,\vec{B}_t
	\end{bmatrix}\begin{bmatrix}
		\vec{\Psi(X)}\\ \vec{U}
	\end{bmatrix}\\
	&\Rightarrow\begin{bmatrix}
		\vec{K}_t,\vec{B}_t
	\end{bmatrix}\approx\vec{\Psi(X')}\begin{bmatrix}
		\vec{\Psi(X)}\\\vec{U}
	\end{bmatrix}^+,
\end{align}
where ${\vec{K}_t\in\mathbb{R}^{N\times N}}$ is the approximated Koopman operator and ${\vec{B}_t \in\mathbb{R}^{N\times p}}$ the lifted input matrix. 
The resulting discrete-time system description for EDMDc prediction is given by
\begin{equation}\label{eq:edmdc_pred}
	\vec{\hat{\Psi}}(\vec{x}_{k+1})=\vec{K}_t\vec{\Psi}(\vec{x}_k)+\vec{B}_t\vec{u}_k.	
\end{equation}
The hat on the symbols emphasizes that the quantities are estimated, cf.~\eqref{eq:edmd}.

\subsection{Stable Koopman Operators}\label{subsec:stable_KO}
\cite{GKS19} established an algorithm that computes a nearby stable discrete-time system to an unstable one. This approach has been applied to the Koopman operator and has shown that the EDMD prediction error drastically reduces when using stable approximated Koopman operators instead of unstable ones (\cite{MAM20}).

The main idea is based on a new characterization for the set of stable matrices: A matrix ${\vec{A}_s\in\mathbb{R}^{n\times n}}$ is stable if there exist ${\vec{S},\vec{O},\vec{T}\in\mathbb{R}^{n\times n}}$ such that ${\vec{A}_s=\vec{S}^{-1}\vec{O}\vec{T}\vec{S}}$ where ${\vec{S}\succ0}$, $\vec{O}$ is orthogonal, ${\vec{T}\succeq0}$ and ${\Vert\vec{T}\Vert\leq1}$. On this basis, the next stable matrix $\vec{A}_s$ to an unstable matrix $\vec{A}$ in terms of the Frobenius norm can then be defined as follows, where the allowed search space is given by the set of stable matrices
\begin{equation}\label{eq:next_stable_matrix}
	\vec{A}_s=\arginf_{\vec{S}\succ0,\vec{O}\in\mathcal{O}(n),\vec{T}\succeq0,\lVert\vec{T}\rVert\leq1} \lVert\vec{A}-\vec{S}^{-1}\vec{O}\vec{T}\vec{S}\rVert_F^2.
\end{equation}
An algorithmic solution for\,\eqref{eq:next_stable_matrix} can be found by a projected gradient descent. More precisely, the matrices ${\vec{S},\vec{O},\vec{T}}$ are initialized in and projected onto the set of feasible matrices in each descent step, where the objective function scores the distance to the original unstable matrix. While projecting onto the set of feasible matrices, the eigenvalues of the matrices are systematically shifted, which is explained in detail below.

Real symmetric matrices $\vec{A}\in\mathbb{R}^{n\times n}$ with eigenvalues ${\lambda_1,\cdots,\lambda_n}$ are diagonalizable by orthogonal matrices ${\vec{V}\in\mathbb{R}^{n\times n}}$ yielding
\begin{equation}
	\vec{A}=\vec{V}\mathrm{diag(\lambda_1,\cdots,\lambda_n)}\vec{V}^\top.
		\label{eq:f_A1}
\end{equation}
Due to this property, we set 
\begin{equation}
	f\left(\vec{A}\right)=\vec{V}\left(\mathrm{diag}(f(\lambda_1),\cdots,f(\lambda_n))\right)\vec{V}^\top,
\end{equation}
where $f$ is any complex-valued function defined on the spectrum of $\vec{A}$. This allows to simply shift the eigenvalues of a real symmetric matrix with the following function
\begin{equation}
	p_{a,b}(\lambda)=\begin{cases}
		a,&\lambda<a\\
		\lambda,&\lambda\in\left[a,b\right]\\
		b,&\lambda>b
	\end{cases}
	\label{eq:f_A2}
\end{equation}
into the interval $\left[a,b\right]$ (\cite{GKS19}). We will use this strategy in Sec.\,\ref{sec:algorithm} to systematically modify the definiteness of a symmetric matrix to satisfy the PCHD conditions (\cite{GS17}). 
\section{Data-Driven PCHD Models}\label{sec:algorithm}
Inspired by the compelling potential of passivity properties and the EDMDc method with subsequently targeted shifting of eigenvalues, we establish a procedure for data-driven models in PCHD form. We do not seek to transform the states into a higher-dimensional space, but to obtain a PCHD model by combining measurement data with prior physical knowledge. The following assumptions are made:
\begin{enumerate}
	\item The necessary condition ${\dim\vec{u}=\dim\vec{y}}$ for the PCHD form is met (see Sec.\,\ref{subsec:pchd}).
	\item Measured or simulated data of $\vec{x}$ and $\vec{u}$ is available in a sufficiently large amount.
	\item Basic physical prior knowledge exists, i.\,e., the energy function of the system.
\end{enumerate}
Again, consider continuous-time control-affine systems\,\eqref{eq:control-affine_system}. The aim is to obtain a data-driven passive continuous-time system, with the PCHD form\,\eqref{eq:pchd} being the preferred choice for this purpose, assuming the following constraints:
\begin{enumerate}
	\item The matrices $\vec{J}$ and $\vec{D}$ are constant. In general, these matrices may depend on $\vec{x}$, so this constraint corresponds to an approximation, which we will analyze later in the application (see Sec.\,\ref{sec:results}). 
	\item During the learning process, $\vec{J}$ and $\vec{D}$ are combined to $\vec{K}=\vec{J}-\vec{D}$. This assumption does not pose a problem because any matrix can be uniquely decomposed into a symmetric and a skew-symmetric matrix. Thus applies 
	\begin{equation}\label{eq:K_decomposition}
		\vec{J}=\tfrac{1}{2}\left(\vec{K}-\vec{K}^\top\right),\vec{D}=-\tfrac{1}{2}\left(\vec{K}+\vec{K}^\top\right).
	\end{equation}
	\item $\vec{B}$ is a constant matrix.
\end{enumerate}
This results in the following system description
\begin{equation}
	\dot{\vec{x}}=\vec{K}\left(\dfrac{\partial V}{\partial\vec{x}}\right)^\top+\vec{B}\vec{u}.
\end{equation}

The measurement data is stacked into
\begin{subequations}\label{eq:data_passive}
\begin{align}
	\label{eq:snapshots_passive}
	\vec{X}&=\begin{bmatrix}
		\vec{x}_1,\vec{x}_2,\cdots,\vec{x}_{M}
	\end{bmatrix}\in \mathbb{R}^{n\times M},\\
	\dot{\vec{X}}&=\begin{bmatrix}
		\dot{\vec{x}}_1,\dot{\vec{x}}_2,\cdots,\dot{\vec{x}}_M
	\end{bmatrix}\in \mathbb{R}^{n\times M},\\
	\vec{U}&=\begin{bmatrix}
		\vec{u}_1,\vec{u}_2,\cdots,\vec{u}_{M}
	\end{bmatrix}\in\mathbb{R}^{p\times M}.
\end{align}
\end{subequations}
Note that unlike EDMDc, see\,\eqref{eq:data_edmdc}, here the time derivatives of $\vec{x}$ are used.

\begin{figure}[t]
	\centering
	\includegraphics[scale=1]{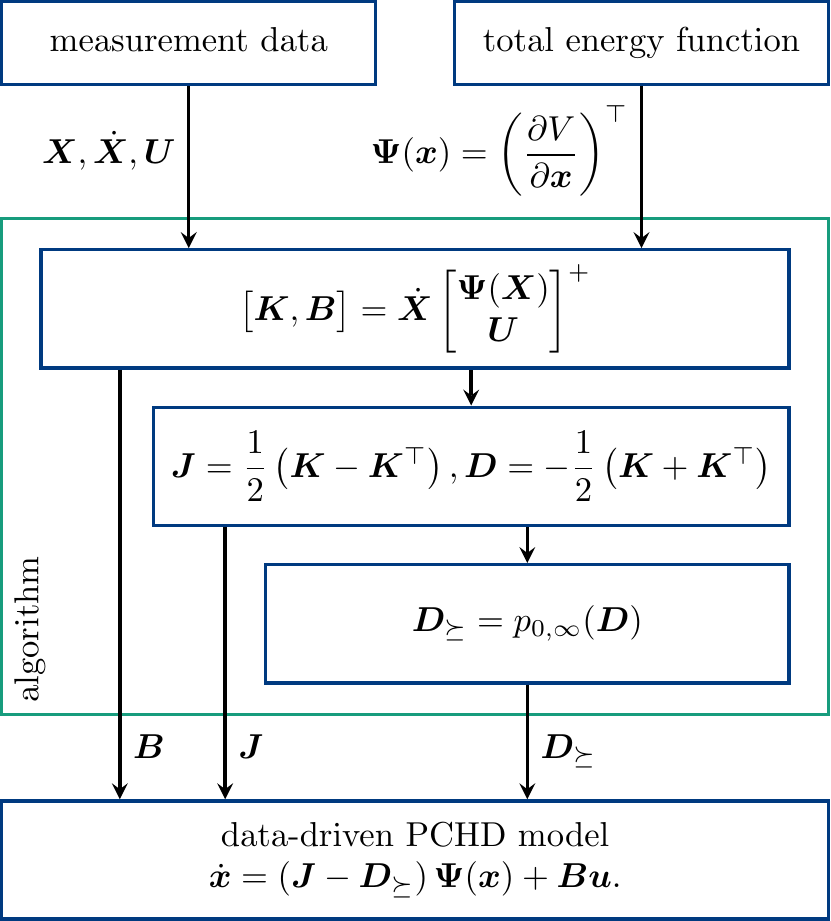}
	\caption{Schematic flow of the algorithm to obtain a data-driven PCHD model.}
	\label{fig:algorithm}
\end{figure}

Prior physical knowledge about the total energy
\begin{equation}
	V(\vec{x})=E_{kin}+E_{pot}
\end{equation}
inside the system is necessarily and used for
\begin{equation}
	\vec{\Psi}(\vec{x})=\left(\dfrac{\partial V}{\partial\vec{x}}\right)^\top,
\end{equation}
yielding
\begin{equation}
	\dot{\vec{x}}=\vec{K}\vec{\Psi}(\vec{x})+\vec{B}\vec{u}.
\end{equation}
The matrices $\vec{K}$ and $\vec{B}$ are approximated over the data, resulting in
 \begin{align}
 		\dot{\vec{X}}&\approx\vec{K}\vec{\Psi}(\vec{X})+\vec{B}\vec{U}=\begin{bmatrix}
 		\vec{K},\vec{B}
 	\end{bmatrix}\begin{bmatrix}
 		\vec{\Psi}(\vec{X})\\\vec{U}
 	\end{bmatrix},\\
 	&\Rightarrow\begin{bmatrix}
 		\vec{K},\vec{B}
 	\end{bmatrix}\approx\dot{\vec{X}}\begin{bmatrix}
 	\vec{\Psi}(\vec{X})\\\vec{U}
 \end{bmatrix}^+.\label{eq:formula1}
 \end{align}
Next, the matrix $\vec{K}$ is decomposed into $\vec{J}$ and $\vec{D}$, cf.\,\eqref{eq:K_decomposition}. To obtain a PCHD form as in\,\eqref{eq:pchd}, it is necessary to ensure that $\vec{D}$ is positive semi-definite. For symmetric matrices, the definiteness follows directly from the properties of the eigenvalues, i.\,e., a symmetric real-valued matrix is positive semi-definite if all its eigenvalues are nonnegative.The eigenvalues of $\vec{D}$ can be set to non-negative using the strategy from Sec.\,\ref{subsec:stable_KO}. According to \eqref{eq:f_A1}-\eqref{eq:f_A2}, a positive semi-definite $\vec{D}$ is thus calculated as follows
\begin{equation}\label{eq:formula2}
	\vec{D}_\succeq=p_{0,\infty}(\vec{D})
\end{equation}
with
\begin{equation}
	p_{0,\infty}(\lambda)=\begin{cases}
		0,&\lambda<0\\
		\lambda,&\lambda\geq0
	\end{cases},
\end{equation}
so that this results in a PCHD system description that meets all criteria and is passive
\begin{equation}
	\dot{\vec{x}}=\left(\vec{J}-\vec{D}_\succeq\right)\vec{\Psi}(\vec{x})+\vec{B}\vec{u}.
\end{equation}
The overall procedure for obtaining a data-driven PCHD model is summarized in Fig. \ref{fig:algorithm}.
\section{Results}\label{sec:results}
The algorithm is demonstrated on two different systems. First, the pendulum is analyzed based on a model, and then we show experimental results on the golf robot.
\subsection{Pendulum}\label{subsec:pendulum}
The proposed framework is simulatively validated using the nonlinear pendulum with damping as an introductory classical control engineering example. Assume the following differential equations
\begin{subequations}
	\begin{align}
		\dot{x}_1&=x_2,\\
		\dot{x}_2&=-\tfrac{g}{l}\sin(x_1)-\tfrac{d}{ml^2}x_2+\tfrac{1}{ml^2}u,
	\end{align}
\end{subequations}
where $x_1$ and $x_2$ are the angle and angular velocity of the pendulum, respectively, and the parameters are assumed to be as shown in Table~\,\ref{tab:parameters_pendulum}. The system input $u$ is the torque on the pendulum. 
\begin{table}[ht]
	\caption{Physical parameters of the pendulum.}
	\label{tab:parameters_pendulum}
	\begin{center}
		\begin{tabular}{cll}
			symbol&physical parameter&value\\
			\hline
			$m$& point mass of the pendulum&  \SI{1}{\kilo\gram}\\ 
			$l$& length of the pendulum&  \SI{0.5}{\meter}\\
			$g$& gravity constant& \SI{9.81}{\meter\per\second^2}\\
			$d$& damper constant& \SI{0.05}{\kilo\gram\meter^2\per\second}\\
			\hline
		\end{tabular}
	\end{center}
\end{table}

Ten~simulated trajectories (numerical integration with RK4 solver) were used to generate the training data. Each trajectory had a duration of \SI{1}{\second}, with a step size of $\Delta t=\SI{0.01}{\second}$ and random initial conditions from the basin of attraction of the stable equilibrium $\vec{x}^*=\begin{bmatrix}
	0,0
\end{bmatrix}^\top$ with piecewise constant $u\in\left[-1,1\right]$.

The total (potential and kinetic) energy is given by
\begin{equation}
	V(\vec{x})=\tfrac{1}{2}ml^2x_2^2+mgl(1-\cos(x_1)),
\end{equation}
yielding
\begin{equation}	
	\vec{\Psi}(\vec{x})=\left(\dfrac{\partial V}{\partial\vec{x}}\right)^\top=\begin{bmatrix}
		mgl\sin(x_1)\\ml^2x_2
	\end{bmatrix}.
\end{equation}
The algorithm presented in Sec.\,\ref{sec:algorithm} returns 
\begin{equation}
	\vec{J}=\begin{bmatrix}
		0&4\\-4&0
	\end{bmatrix},
	\vec{D}=\begin{bmatrix}
		0&0\\0&0.8
	\end{bmatrix},
\vec{b}=\begin{bmatrix}
	0\\4
\end{bmatrix},
\end{equation}
which corresponds to the analytical PCHD model derived from the nonlinear physical model
\begin{equation}
	\resizebox{0.87\hsize}{!}{$
	\vec{J}_{ph}=\begin{bmatrix}
		0&\tfrac{1}{ml^2}\\-\tfrac{1}{ml^2}&0
	\end{bmatrix},
	\vec{D}_{ph}=\begin{bmatrix}
		0&0\\0&\tfrac{d}{m^2l^4}
	\end{bmatrix},\\ \vec{b}_{ph}=\begin{bmatrix}
		0\\\tfrac{1}{ml^2}
	\end{bmatrix}$}.
\end{equation}
$\vec{D}$ is already positive semi-definite and therefore does not need to be modified, so it is $\vec{D}_\succeq=\vec{D}$.

\begin{figure}[ht]
	\centering
	\includegraphics[scale=1]{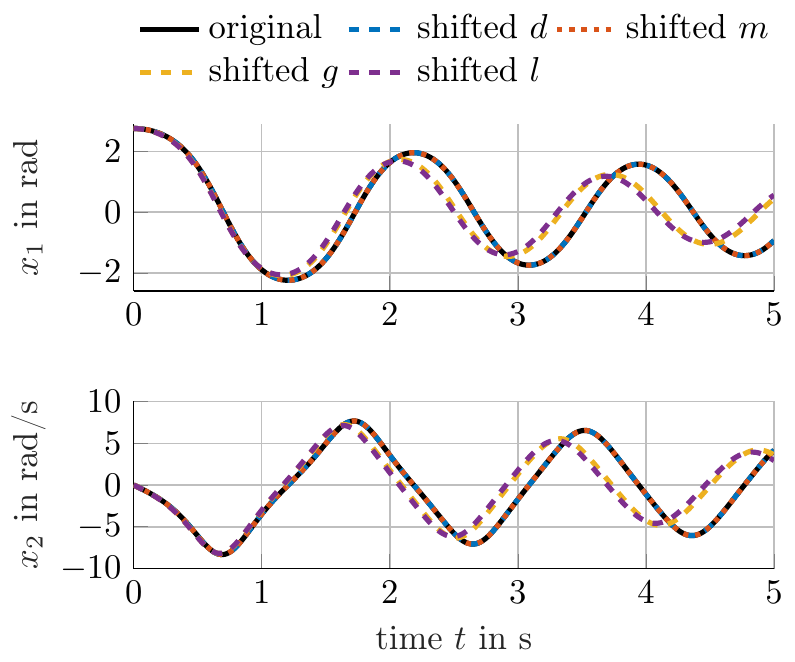}
	\caption{Analysis of possible errors in prior knowledge: Identified data-driven PCHD models for the pendulum with different parameter shifts of the total energy function.}
	\label{fig:pendulum_wrong_parameters}
\end{figure}

Figure \ref{fig:pendulum_wrong_parameters} shows simulation results of the autonomous swinging pendulum, assuming incorrect (shifted) parameters for the total energy function. A deviation of \SI{10}{\percent} from the original value does not pose a problem for the parameters $m$ and $d$, as it is corrected by the algorithm, but leads to poor results for $g$ and $l$. This can be explained by the latter two parameters having a major influence on the oscillation dynamics, i.\,e., the eigenfrequency. Note here that the chosen deviation is just for illustrative purposes; if uncertainty about the system parameters exists, they may generally be identified more accurately i.\,e., optimized for. 
\subsection{Golf Robot}\label{subsec:golf_robot}
\begin{figure}[t]
	\centering
	\includegraphics[width=3.6cm]{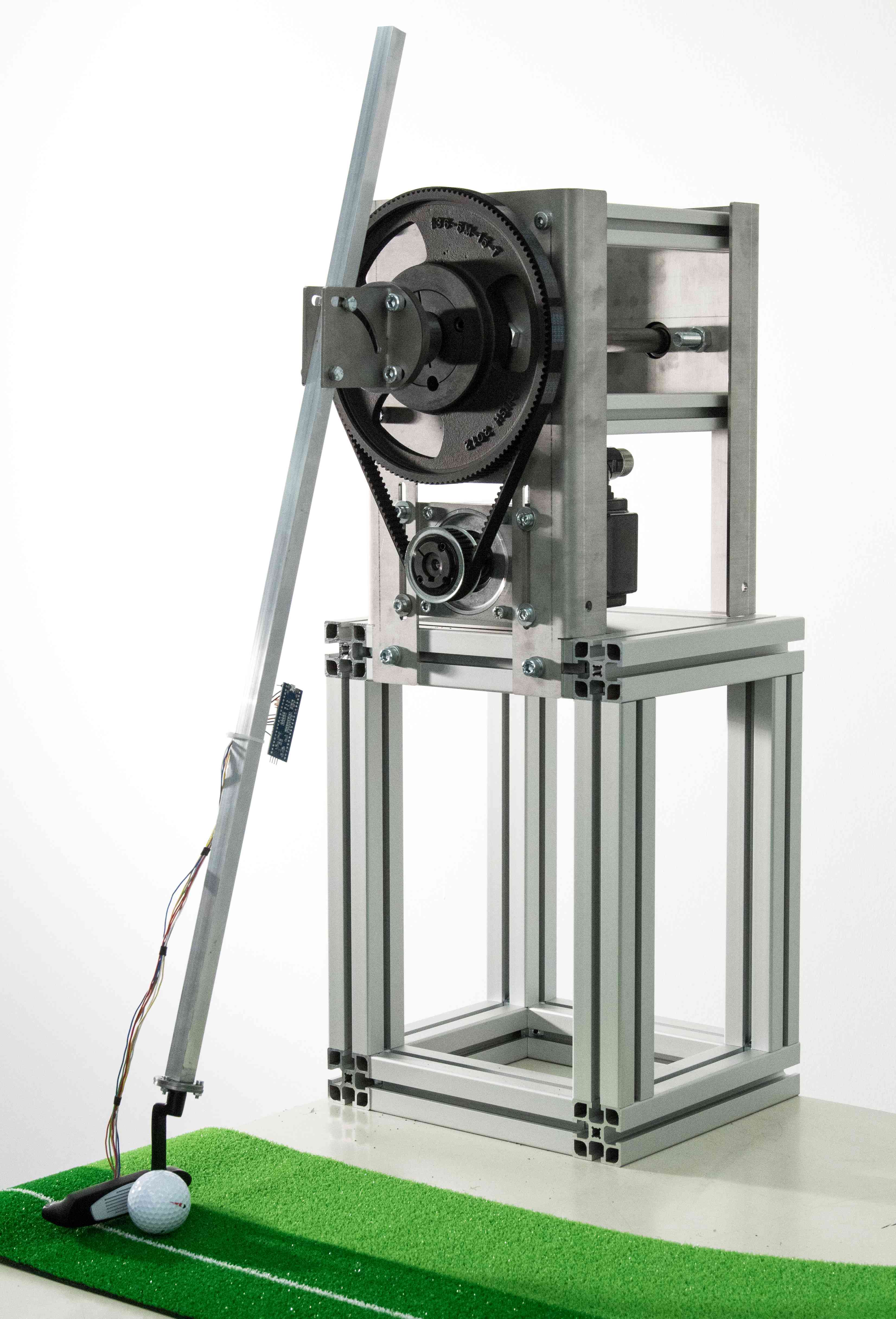}
	\caption{Golf robot as a demonstrator for data-driven methods in control engineering.}
	\label{fig:golfrobot}
\end{figure}

The autonomously putting golf robot shown in Fig.\,\ref{fig:golfrobot} serves as a demonstrator for data-driven methods in control engineering. Two gear shafts connected with a toothed belt drive form the stroke mechanism, where the drive is located on the lower gear shaft and the golf club is mounted on the upper gear shaft.

A simplified physically motivated nonlinear model combines the masses into a single rigid body with torque $u$ as a control input.  The differential equations with parameters shown in Table~\,\ref{tab:parameters_golfrobot} can be described by the following:
\begin{subequations}\label{eq:golfrobot_ode}
	\begin{align}
		\dot{x}_1&=x_2,\\
		\dot{x}_2&=\tfrac{-mga\sin(x_1)-M_d(\vec{x})+4u}{J},
	\end{align} 
\end{subequations}
where $\vec{x}=\begin{bmatrix}
	\varphi,\dot{\varphi}
\end{bmatrix}^T$ contains the angle and angular velocity of the golf club and the nonlinear dissipation torque 
\begin{equation}
	M_d(\vec{x})=dx_2+r\mu\mathrm{sgn}{x_2}\lvert mx_2^2 a+mg\cos{x_1}\rvert
\end{equation}
combines viscous and sliding friction. 

The total energy function is given by
\begin{equation}
	V(\vec{x})=\tfrac{1}{2}Jx_2^2+mga(1-\cos(x_1)),
\end{equation}
yielding
\begin{equation}
	\vec{\Psi}(\vec{x})=\left(\dfrac{\partial V}{\partial\vec{x}}\right)^\top=\begin{bmatrix}
		mga\sin(x_1)\\Jx_2
	\end{bmatrix}.
\end{equation}
At this point, we emphasize that no knowledge of nonlinear dissipation effects is required, neither about the nonlinearities nor about the related parameters.
\begin{table}[b]
	\caption{Physical parameters of the golf robot.}
	\label{tab:parameters_golfrobot}
	\begin{center}
		\begin{tabular}{c p{4.5cm} p{1.85cm}}
			\hline
			symbol&physical parameter&value\\
			\hline
			$m$& mass of the golf club&  \SI{0.5241}{\kilo\gram}\\ 
			$J$& inertia of the rotating mass&  \SI{0.1445}{\kilo\gram\per\meter^2}\\
			$g$& gravity constant& \SI{9.81}{\meter\per\second^2}\\
			$a$& length from the axis of rotation to the center of mass of the golf club& \SI{0.4702}{\meter}\\
			$d$& damper constant& \SI{0.0132}{\kilo\gram\meter^2\per\second}\\
			$r$& length from the axis of rotation to the friction point& \SI{0.0245}{\meter}\\
			$\mu$& coefficient of friction& $1.5136$\\
			\hline
		\end{tabular}
	\end{center}
\end{table}

The training data consists of several test bench measurements with different excitations $u$ of the system (chirp, sine, and step) and a $\SI{1}{\kilo\hertz}$ sampling rate combined into the matrices $\vec{X}$ and $\dot{\vec{X}}$. Because only the output variable $y=x_1=\varphi$ is measured directly, the data for $x_2$ and $\dot{x}_2$ are generated offline by smoothing spline interpolation followed by numerical differentiation. 

The data-driven learning of a PCHD model by\,\eqref{eq:formula1} yields
\begin{equation}\resizebox{0.88\hsize}{!}{$
	\vec{J}=\begin{bmatrix}
		0&6.18\\-6.18&0
	\end{bmatrix}, \vec{D}=\begin{bmatrix}
	0&-0.74\\-0.74&6.44
\end{bmatrix},\vec{b}=\begin{bmatrix}
0\\23
\end{bmatrix},$}
\label{eq:data-driven_model}
\end{equation}
where $\vec{D}$ is not positive semi-definite with eigenvalues ${\lambda_1=-0.08,\lambda_2=6.52}$. Thus, $\vec{D}$ is modified to be positive semi-definite by\,\eqref{eq:formula2} such that
\begin{equation}
	\vec{D}_\succeq=\begin{bmatrix}
		0.08&-0.73\\-0.73&6.44
	\end{bmatrix}
\label{eq:data-driven_pchd_model}
\end{equation}
with eigenvalues $\lambda_1=0,\lambda_2=6.52$.

\begin{figure*}[t]
	\centering
	\includegraphics[scale=1]{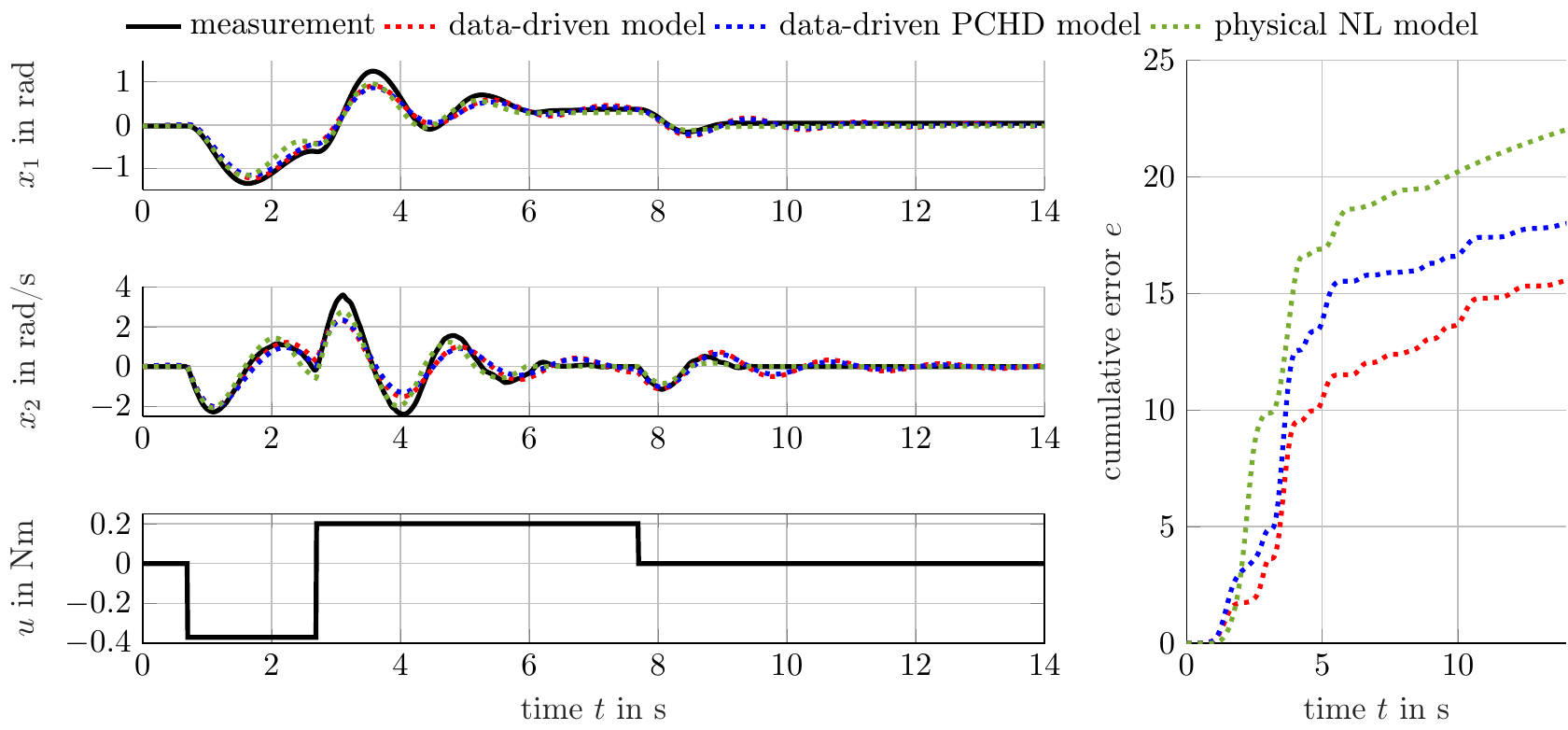}
	\caption{Prediction accuracy based on a test measurement on the golf robot. The simulated data-driven models provide higher prediction accuracy than the simulated classically physics-derived nonlinear model \eqref{eq:golfrobot_ode} (in green) with greatly reduced modeling effort. The PCHD model with positive semi-definite $\vec{D_\succeq}$ \eqref{eq:data-driven_pchd_model} (in blue) provides a slightly lower prediction accuracy than the purely data-driven model \eqref{eq:data-driven_model} (in red).
	}
	\label{fig:golfrobot_passive}
\end{figure*}

To evaluate the model accuracy of the data-driven models, the simulation of a test measurement is compared to the physics-derived model. Figure\,\ref{fig:golfrobot_passive} shows the prediction over time and the cumulative prediction error
\begin{equation}
	e(t_k)=\sum_{m=1}^{k} (x_{1,\text{meas}}(t_m)-x_{1,\text{pred}}(t_m))^2
\end{equation}
of $y=x_1$. Both data-driven models provide higher prediction accuracy with greatly reduced modeling effort than the physics-derived nonlinear model \eqref{eq:golfrobot_ode}. Shifting the eigenvalues from $\vec{D}$ to $\vec{D}_\succeq$ provides a slightly lower prediction accuracy, but exhibits the highly beneficial properties of the PCHD form. 

For comparison, the analytically PCHD model derived from the nonlinear physical model\,\eqref{eq:golfrobot_ode} is given by
\begin{subequations}\label{eq:golfrobot_nl_pchd}
	\begin{align}
		&\vec{J}_{ph} = \begin{bmatrix}
			0&\tfrac{1}{J}\\-\tfrac{1}{J}&0
		\end{bmatrix}\approx\begin{bmatrix}
			0&6.92\\-6.92&0
		\end{bmatrix},\\
		&\vec{D}_{ph}(\vec{x})=\begin{bmatrix}
			0&0\\0&d_{ph}(\vec{x})
		\end{bmatrix},\vec{b}_{ph}=\begin{bmatrix}
		0\\\tfrac{4}{J}
	\end{bmatrix}\approx\begin{bmatrix}
	0\\27.68
	\end{bmatrix}
	\end{align}
\end{subequations}
with
\begin{equation}
	d_{ph}(\vec{x})=\resizebox{0.7\hsize}{!}{$\begin{cases}
			\dfrac{d}{J^2},&x_2=0\\
			\dfrac{d}{J^2}+\dfrac{r\mu}{J^2}\left\lvert\dfrac{mx_2^2a+mg\cos(x_1)}{x_2}\right\rvert,&x_2\neq0	\end{cases}$}.
\end{equation}
Note here that $\vec{D}_{ph}(\vec{x})$ depends on\,$\vec{x}$. Nevertheless, the dominant nonlinearities of the golf robot are likely to be represented by the energy function.
\section{Conclusion \& Outlook}\label{sec:conclusion}
This work has established an algorithm to obtain a PCHD model using measurement data and fundamental physical prior knowledge about the energy stored in the system.
Current research is about designing stabilizing controllers ${\vec{u}=\vec{\beta}(\vec{x})}$ for the data-driven PCHD models by preserving the PCHD structure, so that the closed-loop dynamics is given by
\begin{equation}
	\dot{\vec{x}}=\left(\vec{J}_d(\vec{x})-\vec{D}_d(\vec{x})\right)\left(\dfrac{\partial V_d}{\partial\vec{x}}\right)^\top
\end{equation} 
ensuring stability and robustness features. The desired system behavior is determined by the new energy function $V_d(\vec{x})$, which has a strict local equilibrium at the desired equilibrium $\vec{x}^*$, and the desired interconnection and damping matrices ${\vec{J}_d(\vec{x})=-\vec{J}_d^\top(\vec{x})}$, ${\vec{D}_d(\vec{x})=\vec{D}_d^\top(\vec{x})\succeq0}$, respectively (\cite{OvME02,KL09}). In addition, future research might address how to further extend the algorithm to allow state-dependent matrices for $\vec{J}$, $\vec{D}$, and $\vec{B}$.


\end{document}